\documentclass[format=manuscript, nonacm]{acmart}
\usepackage{caption}
\usepackage{verbatim}
\usepackage{tikz}
\usepackage{xcolor}
\usepackage{bbding} 
\usepackage{microtype}
\usepackage{colortbl}
\usepackage{subcaption}
\usepackage{booktabs} 

\usepackage{titlesec}

\titleformat{\subsubsection} 
  {\normalfont\normalsize\bfseries} 
  {\thesubsection.\arabic{subsubsection}} 
  {1em} 
  {} 

\usepackage{microtype}
\usepackage{graphicx}
\usepackage{multirow}
\usepackage{comment}
\usepackage{booktabs} 
\usepackage{tikz}
\usetikzlibrary{positioning, arrows.meta, calc}

\renewcommand\footnotetextcopyrightpermission[1]{} 
\AtBeginDocument{%
  }
    
\setcopyright{acmlicensed}
\copyrightyear{2018}
\acmYear{2018}
\acmDOI{XXXXXXX.XXXXXXX}
\acmConference[Conference acronym 'XX]{Make sure to enter the correct
  conference title from your rights confirmation email}{June 03--05,
  2018}{Woodstock, NY}

\acmISBN{978-1-4503-XXXX-X/2018/06}

\author{Fangming Cui}
\affiliation{%
  \institution{Beijing}
  \city{Beijing}
  \country{China}
  }

\author{Sunan Li}
\affiliation{%
  \institution{Shanghai}
  \city{Beijing}
  \country{China}
  }
  
\author{Jiahong Li}
\affiliation{%
  \institution{Shanghai}
  \city{Beijing}
  \country{China}
  }

\begin{document}

\title{A Brief Overview: On-Policy Self-Distillation In Large Language Models}



\begin{abstract}
On-Policy Self-Distillation (OPSD) is a unified learning framework in which a single large language model acts simultaneously as both teacher and student. Unlike conventional knowledge distillation that relies on a separate, often larger teacher model, OPSD operates under different contextual roles: the teacher policy is granted privileged access to verified reasoning traces, while the student policy observes only the problem statement. OPSD is trained to minimize per‑token distributional divergence between the two roles over trajectories sampled from the student itself, thereby aligning its own reasoning behavior with solution‑aware rationalizations. OPSD eliminates the need for an external teacher, directly leverages ground‑truth solution information, and resolves the distribution mismatch inherent in off‑policy distillation. 
OPSD typically reduces GPU memory consumption by approximately $40\%$-$60\%$ compared to standard On-Policy Distillation (OPD). 
In this paper, we present a brief analysis of the conceptual foundations, methodological innovations, and principled designs underlying recent advances in OPSD for large language models. 
This discussion, crafted from the perspective of beginners in this field, aims to provide a concise overview of the design principles and emerging patterns of OPSD in LLMs, intended for researchers who are similarly new to this area.

\end{abstract}



\maketitle

\makeatletter
\definecolor{forcePurple}{RGB}{128, 0, 128} 

\let\oldcite\cite

\renewcommand{\cite}[1]{%
  \textcolor{forcePurple}{\oldcite{#1}}%
}

\ifdefined\citep
  \let\oldcitep\citep
  \renewcommand{\citep}[1]{\textcolor{forcePurple}{\oldcitep{#1}}}
\fi
\ifdefined\citet
  \let\oldcitet\citet
  \renewcommand{\citet}[1]{\textcolor{forcePurple}{\oldcitet{#1}}}
\fi

\makeatother

\section{Introduction}
The "Three Board Axe" for post-training of large language models is: SFT (Supervised Fine-Tuning), RL (Reinforcement Learning), and OPD (On-Policy Distillation).
SFT (Supervised Fine-Tuning) involves directly mimicking expert trajectories, which is straightforward and efficient~\cite{sft1,sft2,sft3,sft4}. However, it is inherently off‑policy, there is a mismatch between the training distribution and the inference distribution. This misalignment leads to poor generalization and, during continual learning, often results in catastrophic forgetting.
RL (Reinforcement Learning) method employs verifiable rewards for on-policy training, thereby addressing distribution shift~\cite{shao2022deepseek,zheng2025group,yu2025dapo,sapo}. However, the reward signal is extremely sparse, a single scalar uniformly assigned to all tokens-making credit assignment a critical bottleneck. Moreover, when all rollouts in a batch are either entirely correct or entirely incorrect, the gradient signal vanishes completely.
OPD (On-Policy Distillation) provides dense token-level supervision but relies on a stronger external teacher model~\cite{opd,opd2}. Additionally, it must contend with the difficulty of learning when the teacher model is too powerful and differs too significantly from the student model. Moreover, the additional deployment of a teacher model imposes excessive demand on GPU memory.
Currently, there exists a solution that achieves both on‑policy training and dense supervision without relying on an external teacher: let the same model play two roles. The student sees only the problem, while the teacher additionally receives privileged information (e.g., ground‑truth solutions or reasoning traces). Both roles share the same model parameters but are conditioned on different contexts. During training, the model minimizes the divergence between the two distributions over trajectories sampled by the student itself. This solution is called On-Policy Self-Distillation (OPSD)~\cite{Self-Distilled}.
Prior surveys have provided comprehensive perspective of On-Policy Distillation (OPD) in large language models. Work~\cite{Survey2} offers the phenomenology, mechanism, and recipe.  Work~\cite{Survey} targets the landscape of OPD.

\begin{table}[htbp]
\centering
\caption{Comparison of Different Methods}
\label{tab:method_comparison}
\begin{tabular}{@{}llll@{}}
\toprule
Method  & Supervision & Key Strength & Key Weakness \\
\midrule
SFT  &  Sparse (Sequence) & Simple, stable, fast & Exposure bias, no exploration \\
GRPO &  Sparse (Scalar Reward) & No value model, on-policy RL & Credit assignment problem, high variance \\
OPD  &  Dense (Token-level) & Stability, fine-grained learning & Requires strong external teacher model \\
OPSD &  Dense (Token-level) & No external teacher, self-contained & Requires careful handling of privileged info \\
\bottomrule
\end{tabular}
\end{table}

\textbf{Our Contributions.} In this paper, we take a focused look at the mathematical formulations, implementation insights, and technical innovations in On‑Policy Self‑Distillation (OPSD) for large language models. We present the material in a concise, digestible format to help readers quickly grasp the mechanics and advantages of the OPSD framework.

\section{Background}
In this section, we introduce the principles of the classic SFT, GRPO (RL method), OPD, and OPSD.

\subsection{Supervised Fine-Tuning (SFT)} 
SFT is the foundational baseline for adapting pre-trained LLMs to downstream tasks. It operates on a simple principle: maximum likelihood estimation (MLE) on expert demonstrations. Given a dataset $\mathcal{D} = \{(x^{(i)}), y^{(i)}\}$ of prompt-answer pairs, SFT trains the policy $\pi_{\theta}$ to maximize the probability of the expert's next token. It is inherently off-policy—the training distribution (expert trajectories) is disjoint from the inference distribution (model samples), leading to exposure bias.
The standard SFT objective is the negative log-likelihood (NLL) loss:

\begin{equation}
    \begin{aligned}
\mathcal{L}_{\text{SFT}}(\theta) = -\mathbb{E}_{(x,y) \sim \mathcal{D}} \left[ \sum_{t=1}^{y } \log \pi_\theta(y_t
 x, y_{<t}) \right],
 \end{aligned}
\end{equation}
where $\pi_{\theta}$: The LLM policy (probability distribution over tokens).
$x$: Input prompt/context.
$y = (y_1, \dots, y_T)$: Target response sequence.
$y_{<t}$: All tokens preceding position $t$.
To prevent catastrophic forgetting of the pre-training distribution, a KL penalty is often added, anchoring the policy to the initial pre-trained model $\pi_{\text{ref}}$:

\begin{equation}
    \begin{aligned}
\mathcal{L}_{\text{SFT+KL}}(\theta) = \mathcal{L}_{\text{SFT}}(\theta) + \beta \cdot \mathbb{E}_{x \sim \mathcal{D}} \left[ \text{KL}\left( \pi_{\text{ref}}(\cdot x) \parallel \pi_\theta(\cdot
 x) \right) \right],
\end{aligned}
\end{equation}
where $\beta$ is a regularization coefficient.
However, this method has two limitations: 1) Distributional Shift (Exposure Bias): During training, the model conditions on ground-truth tokens $y_{<t}$; at inference, it conditions on its own (potentially erroneous) samples $\hat{y}_{<t}$. This mismatch causes error accumulation.
2) Lack of Exploration: SFT merely imitates; it cannot learn from or improve upon the demonstration data.

\subsection{GRPO}
Developed by DeepSeek, GRPO~\cite{shao2022deepseek} is a cutting-edge reinforcement learning algorithm that improves the reasoning performance of large language models (LLMs). 
GRPO's key methodological innovation is to forgo the resource-intensive Critic model. It accomplishes this by sampling a group of responses for each prompt and computing a relative advantage signal through within-group reward normalization, eliminating the need for absolute value estimation.
By employing a "group-relative" advantage signal, GRPO yields significant reductions in both memory footprint and compute overhead. 
Following the generation of $G$ responses per query $x$, GRPO derives two key quantities per token $y_{i,t}$: the importance ratio $w_{i,t}(\theta)$ and the advantage estimate $\widehat{A}_{i,t}$. The latter is defined at the response level, meaning the same value $\widehat{A}_{i}$ applies to every token within a given output sequence $y_{i}$.
The GRPO loss is defined as:
\begin{equation}
\begin{aligned}
&\mathcal{J}_{\mathrm{GRPO}}(\theta)
=
\mathbb{E}_{x \sim \mathcal{D},\;
\{y_i\}_{i=1}^G \sim \pi_{\theta_{\text{old}}}(\cdot|x)}
\\
&\qquad\qquad
\bigg[
\frac{1}{G}
\sum_{i=1}^{G}
\frac{1}{|y_i|}
\sum_{t=1}^{|y_i|}
\min \Big(
w_{i,t}(\theta)\, \widehat{A}_{i,t},
\mathrm{clip}\big(w_{i,t}(\theta), 1-\varepsilon, 1+\varepsilon\big)\, \widehat{A}_{i,t}
\Big)
\bigg],
\end{aligned}
\end{equation}
where
\begin{equation}
    \begin{aligned}
w_{i, t}(\theta)=\frac{\pi_{\theta}\left(y_{i, t} \mid x, y_{i,<t}\right)}{\pi_{\theta_{\text {old }}}\left(y_{i, t} \mid x, y_{i,<t}\right)}, \quad \widehat{A}_{i, t}=\widehat{A}_{i}=\frac{r\left(x, y_{i}\right)-\text { mean }\left(\left\{r\left(x, y_{i}\right)\right\}_{i=1}^{G}\right)}{\operatorname{std}\left(\left\{r\left(x, y_{i}\right)\right\}_{i=1}^{G}\right)}.
    \end{aligned}
\end{equation}
However, this method has two limitations: 1) Sparse Credit Assignment: Despite being on-policy, the reward is typically a single scalar for the entire sequence (e.g., 1 for correct final answer, 0 otherwise). This makes it extremely difficult to attribute credit to specific tokens (the credit assignment problem). 2) High Variance: If all responses in a group are correct (or all incorrect), the advantage signal collapses to zero, leading to vanishing gradients.


\subsection{On-Policy Distillation (OPD)}

OPD, formalized by the Generalized Knowledge Distillation (GKD) framework, bridges the gap between RL and imitation learning by providing dense, token-level supervision on the student's own trajectories.
Instead of learning from a fixed dataset (off-policy), the student policy $\pi_\theta$ generates responses on-policy (by sampling or beam search). A powerful, frozen teacher model $\pi_{\text{teacher}}$ then provides a probability distribution over each token in these generated trajectories. The student minimizes the divergence (typically reverse KL) between its output and the teacher's at every token position.
The OPD objective under the GKD framework is:

\begin{equation}
    \begin{aligned}
\mathcal{L}_{\text{OPD}}(\theta) = \mathbb{E}_{x \sim \mathcal{D}, y \sim \pi_\theta(\cdot| x)} \left[ \sum_{t=1}^{ \lvert y \rvert } \text{KL}\left( \pi_\theta(\cdot |x, y_{<t}) \parallel \pi_{\text{teacher}}(\cdot
 |x, y_{<t}) \right) \right].
\end{aligned}
\end{equation}
$y \sim \pi_\theta(\cdot | x)$: A response sampled from the student's current policy (on-policy).
$\text{KL}(p \parallel q)$: Reverse KL divergence, which is mode-seeking and avoids averaging over multiple teacher modes.
Reverse KL $(\text{KL}(p \parallel q)$ penalizes the student for assigning probability to tokens the teacher considers unlikely. This forces the student to commit to a single high-quality mode of the teacher's distribution, producing sharper, more deterministic policies.
However, this method has two limitations: 1) Teacher Dependency: Requires a separate, significantly stronger teacher model, which increases memory footprint and computational cost.
2) Capacity Gap: If the student is too weak relative to the teacher, it may fail to approximate the teacher's complex distribution, leading to optimization instability.

\begin{figure*}[t]
\centerline{\includegraphics[width=1.03\columnwidth]{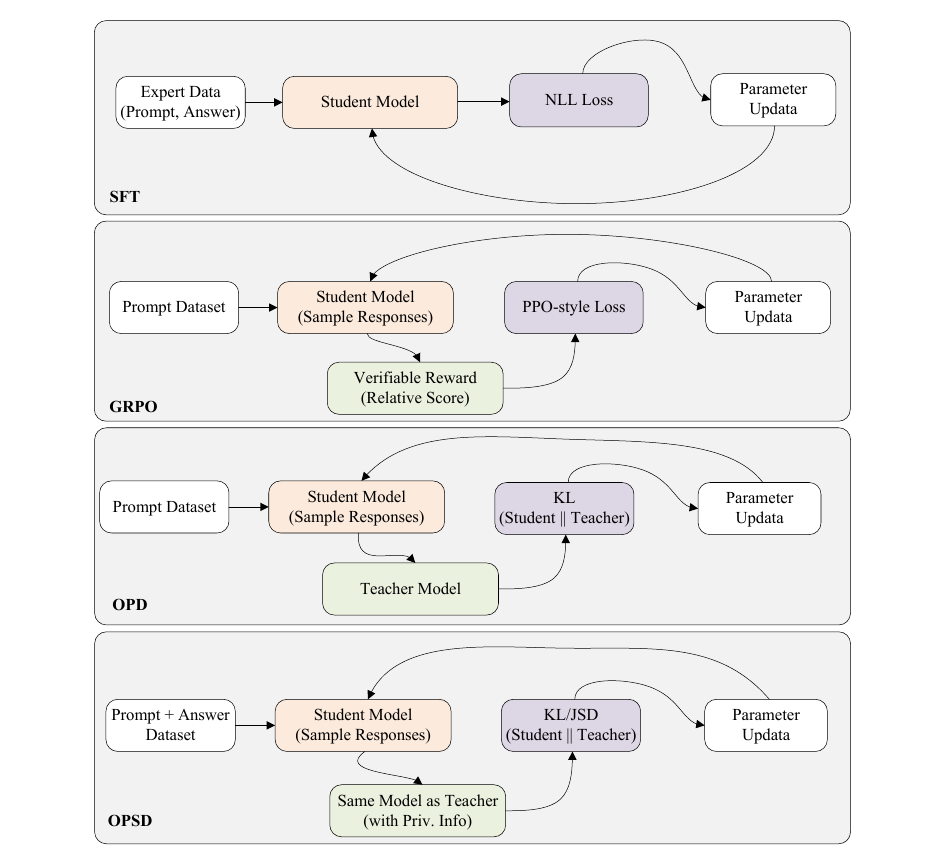}}
\caption{Demonstration of SFT, GRPO, OPD and OPSD.}
\label{rader}
\end{figure*} 
\subsection{On-Policy Self-Distillation (OPSD)}

OPSD is an elegant extension of OPD that eliminates the external teacher by having the same model act as both student and teacher, using privileged information to break the symmetry.
The core hypothesis is that evaluation is easier than generation. A model, when provided with the correct answer (privileged information), can generate a superior reasoning trace. In OPSD, the model is split into two roles with shared parameters $\theta$.
Student $(\pi_{\theta}^{\text{stud}})$: Conditions only on the prompt $x$.
Teacher $(\pi_{\theta}^{\text{teach}})$: Conditions on the prompt $x$ and the ground-truth answer $y^*$ (or other privileged info).
The student generates a rollout $y \sim \pi_{\theta}^{\text{stud}}(\cdot | x)$. The teacher then provides token-level supervision for this specific rollout, and the student minimizes the divergence between the two distributions.

\begin{equation}
    \begin{aligned}
\mathcal{L}_{\text{OPSD}}(\theta) = \mathbb{E}_{(x,y^*) \sim \mathcal{D}, y \sim \pi_{\theta}^{\text{stud}}(\cdot |x)} \left[ \sum_{t=1}^{ \lvert y \rvert } \text{KL}\left( \pi_{\theta}^{\text{stud}}(\cdot |x, y_{<t}) \parallel \pi_{\theta}^{\text{teach}}(\cdot
 |x, y^*, y_{<t}) \right) \right].
\end{aligned}
\end{equation}
Privileged Information $(y^*)$: This is the "answer key" given only to the teacher. It allows the teacher to generate a corrected or verified reasoning path that the student must align with.
Shared Weights: Since $\theta$ is shared, the model is effectively learning to make its "blind" (student) behavior resemble its "informed" (teacher) behavior.
In the OPSD (On-Policy Self‑Distillation) framework, the per‑token divergence between the teacher and student distributions can be measured by a variety of divergence metrics. Among them, a generalized Jensen‑Shannon divergence (JSD) is often mentioned as a stable, symmetric alternative to standard KL divergence.
Given a weighting parameter  $\beta \in [0,1]$ , the generalized JSD between two distributions  $p_T$  (teacher) and  $p_S$  (student) is defined as:

\begin{equation}
\begin{aligned}
\text{JSD}\_\beta(p_T \mid\mid p_S) = \beta \, D_{\text{KL}}\big(p_T \mid\mid m \big) + (1-\beta) \, D_{\text{KL}}\big(p_S \mid\mid m \big),
\end{aligned}
\end{equation}
where  $m$  is the mixture distribution formed by a convex combination of the two:

\begin{equation}
    \begin{aligned}
m = \beta \, p_T + (1-\beta) \, p_S,
\end{aligned}
\end{equation}
and $ D_{\text{KL}}(P \| Q) $ denotes the standard Kullback–Leibler divergence.
Unlike KL divergence, JSD is symmetric $(\text{JSD}_{\beta}(p_T \parallel p_S) = \text{JSD}_{1-\beta}(p_S \parallel p_T))$. This is natural in a self‑distillation setting where teacher and student are the same model under different conditioning.
The standard JSD (with  $\beta$ = 0.5 ) is bounded between $0$ and  $\log 2$ , which helps stabilize training and prevents gradient explosion.
The mixture distribution  m  acts as a “smoothing bridge” between teacher and student, making the loss landscape easier to optimize.
Incorporating JSD into the OPSD Objective
When JSD is chosen as the divergence  $D$  in the OPSD loss, the per‑token training objective becomes:

\begin{equation}
\begin{aligned}
\mathcal{L}_{\text{OPSD-JSD}}(\theta) = \mathbb{E}_{(x,y^*)\sim \mathcal{D}} \mathbb{E}_{\hat{y} \sim p_S(\cdot |x)} \Bigg[ \frac{1}{\hat{y}} \sum_{t=1}^{\hat{y} } \text{JSD}_\beta\Big(p_T(\cdot| x, y^*, \hat{y}_{<t}) \,,\big( p_S(\cdot |x, \hat{y}_{<t}) \Big) \Bigg].
\end{aligned}
\end{equation}
$\beta = 0.5$  gives the classic, symmetric Jensen‑Shannon divergence. $\beta \neq 0.5$  allows the objective to be biased toward either the teacher ( $\beta > 0.5$ ) or the student ( $\beta < 0.5$ ), which can be useful for controlling the “strength” of teacher supervision.
When  $\beta \to 1$ ,  $\text{JSD}_\beta$  approaches the forward KL $ D_{\text{KL}}(p_T \| p_S) $.
When  $\beta \to 0$ , it approaches the reverse KL $D_{\text{KL}}(p_S \| p_T)$.
The generalized JSD can be seen as a smooth interpolation between forward and reverse KL, offering a tunable trade‑off between mode‑covering and mode‑seeking behavior.

\section{Designs}

\subsection{Hybrid Policy Optimization}

Core Idea: Move beyond pure RL (sparse rewards) or SL (supervised learning) by blending RL's advantage function with self-distillation's logit signals~\cite{Self-Distilled,Privileged,On-policy,Online}. RL provides the "direction," while SD provides the "magnitude."

RLSD~\cite{Self-Distilled-RLVR}: Combines RLVR with self-distillation to prevent information leakage. SRPO~\cite{Unifying}: Dynamic routing, SRPO corrects samples processed by RL, failed samples by SD.
GEAR~\cite{GEAR}: Uses divergence signals to identify semantic deviation points in trajectories for advantage function segmentation.
VISD~\cite{VISD}: VISD introduces structured self-distillation for video reasoning, using a video-aware judge to decompose reasoning quality into multiple dimensions and provide diagnostically meaningful feedback for token-level supervision.
RLRT~\cite{Rebellious}: 
This is interpreted as a new form of exploration in RLVR: not uniform diversity, but exploration that leverages the student’s own successful trajectories.
HDPO~\cite{HDPO}: HDPO targets failure modes by generating privileged rollouts with ground-truth info, distilling the teacher's token-level distribution to augment standard RL. The shared weights ensure a bounded realizability gap.

\subsection{Feedback-Driven Distillation}

Core Idea: Leverage environmental feedback (e.g., compiler errors, failed unit tests, human critique) as a "reflection" context~\cite{Pi-Play,Learnwhereto,Multilingual}. The model regenerates a corrected output, which is then distilled back into itself. 

SDPO~\cite{Reinforcement}: SDPO improves sample efficiency and final accuracy across reasoning and code tasks, and can accelerate test-time discovery.
Skill-SD~\cite{Skill-SD}: Skill-SD distills an agent’s own trajectories into natural language skills as privileged teacher context, enabling dense supervision via an importance-weighted reverse-KL loss for stable training.
GATES~\cite{GATES}: Instead of assuming tutor correctness, GATES derives supervision from tutor consensus via multiple sampled reasoning traces. 
PAINT~\cite{PAINT}: PAINT introduces adaptive teacher exposure: it masks the verified solution based on rollout-reference overlap and applies energy-space interpolation on entropy-mismatch tokens.
SD-Zero~\cite{Self-Distillation-Zero}: SD-Zero uses a single model as both Generator and Reviser. The Reviser conditions on the Generator’s output and a binary reward to improve it. 

\subsection{Training Stability \ Calibration}

Core Idea: The primary risk of self-distillation is signal degeneration or overconfidence from "teaching itself." 

AntiSD~\cite{Anti-Self-Distillation}: In mathematical reasoning, AntiSD reverses self-distillation: it maximizes teacher-student divergence and uses an entropy gate, providing a bounded, drop-in alternative to standard distillation.
CREDIT~\cite{From}: Analyzes rewards from a pointwise mutual information (PMI) perspective, filtering out input-agnostic shortcut signals.
PACED~\cite{PACED}: Standard distillation wastes compute. PACED weights problems by the student's pass rate, concentrating training on the zone of proximal development. It requires no extra parameters.
ATESD~\cite{Adaptive}: Teacher exposure is critical for self-distillation. ATESD makes it a learnable variable, using a Beta-policy controller and a delayed reward based on the student's future improvement, rather than a fixed hyperparameter.
EGRSD~\cite{Respecting}: EGRSD integrates a reward-grounded direction, a teacher–student likelihood-ratio magnitude, and a novel teacher-entropy confidence gate. 
OGLS-SD~\cite{OGLS-SD}: 
It utilizes verifiable outcome rewards to compare successful against unsuccessful on-policy trajectories, thereby calibrating the logits produced by the teacher model.
Work~\cite{why} analyzes the side effect of self-distillation where the teacher's "rich conditioning" suppresses the model's expression of uncertainty, leading to in-domain improvement but OOD performance degradation. 

\subsection{Reasoning Compression \ Acceleration}

Core Idea: The goal is not to improve accuracy, but to accelerate inference. By teaching the model to generate more concise chains-of-thought (CoT) or reduce decoding steps, it achieves "model slimming."

CRISP~\cite{On-Policy-Self}: Distills from a teacher prompted to "be concise."
Multi-Token~\cite{Multi-Token}: Distills a next-token predictor into a multi-token predictor.
D-OPSD~\cite{OPSDL}: For diffusion models, preserves "few-step inference" capability during fine-tuning.

\subsection{Capability Internalization \ Transfer}

Core Idea: Distill capabilities demonstrated under specific configurations (e.g., with external toolchains, in high-resource languages, with long context) into the base model, enabling it to perform well under standard settings~\cite{Crosslingual,Self-Distillation,Healthcare}.

OPHSD~\cite{Training}: OPHSD uses the harness-augmented model as a teacher to distill its capabilities into a standalone student, achieving robust generalizability.
MSD~\cite{Multilingual}: Transfers safety alignment capabilities from high-resource to low-resource languages.
OPSDL~\cite{OPSDL}: Uses the model's own short-context capability to supervise its long-context generation, reducing hallucinations.
IRIS~\cite{IRIS}: IRIS framework provides a unified theoretical perspective that optimizes the learning dynamics of a model utilizing its own generated content. This process essentially internalizes the knowledge produced by the model itself into stronger generalization abilities. 
Work~\cite{Embarrassingly} proposes "simple self-distillation (SSD)," which improves code generation by sampling from the model's own outputs and fine-tuning on them. 

\subsection{Vision \ Multimodal Distillation}

Core Idea: Apply self-distillation paradigms to visual generation (diffusion models) or multimodal understanding, focusing on feature alignment and preserving few-step reasoning.

D-OPSD~\cite{D-OPSD}: Fine-tuning few-step diffusion models risks damaging their speed. D-OPSD enables on-policy self-distillation: the student generates from text, while the teacher, seeing the text and target image, provides supervision. This preserves the model's efficient inference capability.
RLSD~\cite{Self-Distilled-RLVR}: RLSD hybridizes RLVR and OPSD, using RLVR for direction and self-distillation for magnitude, achieving higher convergence and training stability.

\subsection{Structured Trajectory Distillation}

Core Idea: Move beyond simple per-token imitation by structurally processing reasoning trajectories—through segmentation, re-calibration, or abstraction—to align distillation with semantic units~\cite{VISD}.

GEAR~\cite{GEAR}: Groups trajectories into adaptive segments based on "semantic deviation points."
TABOM~\cite{Self-Distilled-Trajectory}: Replays and abstracts historical trajectories to distill key decision points. It models the denoising order via a Boltzmann distribution and uses a pairwise ranking loss to align training with the model's easy-to-hard inference certainty.

\subsection{Fundamental Paradigms \ Theory}

Core Idea: Investigate the underlying theory of self-distillation, such as generalization bounds, training dynamics, and reward formation mechanisms, providing a foundation for practice.

UniSD~\cite{UniSD}: Generalization analysis from a unified perspective.
CREDIT~\cite{From}: Information-theoretic decomposition of self-distillation rewards.
PBSD~\cite{Preference}: Theoretical framework for preference-based self-distillation.

\section{Settings}
Experiments are allowed to be conducted on 8 A100 GPUs, which can cover most scenarios. Therefore, it is conveniently to research institutes and units with limited training resources.
OPSD typically reduces GPU memory consumption by approximately $40\%$-$60\%$ compared to standard OPD. In extreme large-model scenarios, OPSD is often the only feasible option, whereas OPD would lead directly to an out-of-memory error. 

\textbf{Mathematical Reasoning.}
GSM8K: Grade-school math word problems, testing basic multi-step reasoning~\cite{gsm8k}. 
MATH: Competition-level, high-difficulty math (AMC/AIME), testing extreme capability~\cite{math}. 
AIME: Very high-difficulty integer/geometry competition problems~\cite{aime}. OpenThoughts: The value of OpenThoughts lies in clarifying the teacher‑model selection principle and the efficacy of data synthesis~\cite{OpenThoughts}.
\textbf{Code Generation.}
HumanEval: 164 hand-written Python function completion problems~\cite{HumanEval}. 
MBPP: 1k entry-level Python programming problems from Google~\cite{mbpp}. 
LiveCodeBench: Continuously updated code evaluation benchmark (includes competition problems)~\cite{livecodebench}. 
\textbf{Agent \ Interactive Environments.}
WebShop: Simulated online shopping (HTML interaction)~\cite{Webshop}. 
ALFWorld: Text-based game (household tasks)~\cite{ALFWorld}. 
\textbf{General \ Safety Alignment.}
UltraFeedback: Large-scale, multi-dimensional preference data~\cite{UltraFeedback}. 
BeaverTails: Safety alignment and red-teaming dataset~\cite{BeaverTails}. 
\textbf{Vision \ Multimodal.}
COCO: Image generation and captioning~\cite{coco}. 
MMMU: Multidisciplinary visual question answering~\cite{MMMU}.

\section{Conclusion} 
OPSD has advantages in training costs and is very suitable for the industry's rapid attempts to empower business needs. This paper offers a brief looking of emerging On‑Policy Self‑Distillation (OPSD) in large language models. It investigates how foundation models function dually as driving forces and computational substrates for sophisticated OPSD systems. This looking also organizes the experimental GPU memory resources and commonly used datasets. 
We hope this paper offers researchers a  concise looking of the current frontier, serving as a trying for future efforts to designs.

\bibliographystyle{ACM-Reference-Format}
\bibliography{sample-base}


\begin{thebibliography}{63}


\ifx \showCODEN    \undefined \def \showCODEN     #1{\unskip}     \fi
\ifx \showISBNx    \undefined \def \showISBNx     #1{\unskip}     \fi
\ifx \showISBNxiii \undefined \def \showISBNxiii  #1{\unskip}     \fi
\ifx \showISSN     \undefined \def \showISSN      #1{\unskip}     \fi
\ifx \showLCCN     \undefined \def \showLCCN      #1{\unskip}     \fi
\ifx \shownote     \undefined \def \shownote      #1{#1}          \fi
\ifx \showarticletitle \undefined \def \showarticletitle #1{#1}   \fi
\ifx \showURL      \undefined \def \showURL       {\relax}        \fi
\providecommand\bibfield[2]{#2}
\providecommand\bibinfo[2]{#2}
\providecommand\natexlab[1]{#1}
\providecommand\showeprint[2][]{arXiv:#2}

\bibitem[Agarwal et~al\mbox{.}(2024)]%
        {opd}
\bibfield{author}{\bibinfo{person}{Rishabh Agarwal}, \bibinfo{person}{Nino Vieillard}, \bibinfo{person}{Yongchao Zhou}, \bibinfo{person}{Piotr Stanczyk}, \bibinfo{person}{Sabela Ramos}, \bibinfo{person}{Matthieu Geist}, {and} \bibinfo{person}{Olivier Bachem}.} \bibinfo{year}{2024}\natexlab{}.
\newblock \bibinfo{title}{On-Policy Distillation of Language Models: Learning from Self-Generated Mistakes}.
\newblock
\showeprint[arxiv]{2306.13649}~[cs.LG]
\urldef\tempurl%
\url{https://arxiv.org/abs/2306.13649}
\showURL{%
\tempurl}


\bibitem[Austin et~al\mbox{.}(2021)]%
        {mbpp}
\bibfield{author}{\bibinfo{person}{Jacob Austin}, \bibinfo{person}{Augustus Odena}, \bibinfo{person}{Maxwell Nye}, \bibinfo{person}{Maarten Bosma}, \bibinfo{person}{Henryk Michalewski}, \bibinfo{person}{David Dohan}, \bibinfo{person}{Ellen Jiang}, \bibinfo{person}{Carrie Cai}, \bibinfo{person}{Michael Terry}, \bibinfo{person}{Quoc Le}, {et~al\mbox{.}}} \bibinfo{year}{2021}\natexlab{}.
\newblock \showarticletitle{Program Synthesis with Large Language Models}.
\newblock \bibinfo{journal}{\emph{arXiv preprint arXiv:2108.07732}} (\bibinfo{year}{2021}).
\newblock


\bibitem[Chen et~al\mbox{.}(2026)]%
        {Self-Distilled-Trajectory}
\bibfield{author}{\bibinfo{person}{Kecheng Chen}, \bibinfo{person}{Ziru Liu}, \bibinfo{person}{Xijia Tao}, \bibinfo{person}{Hui Liu}, \bibinfo{person}{Yibing Liu}, \bibinfo{person}{Xinyu Fu}, \bibinfo{person}{Shi Wu}, \bibinfo{person}{Suiyun Zhang}, \bibinfo{person}{Dandan Tu}, \bibinfo{person}{Lingpeng Kong}, \bibinfo{person}{Rui Liu}, {and} \bibinfo{person}{Haoliang Li}.} \bibinfo{year}{2026}\natexlab{}.
\newblock \bibinfo{title}{Self-Distilled Trajectory-Aware Boltzmann Modeling: Bridging the Training-Inference Discrepancy in Diffusion Language Models}.
\newblock
\showeprint[arxiv]{2605.11854}~[cs.CL]
\urldef\tempurl%
\url{https://arxiv.org/abs/2605.11854}
\showURL{%
\tempurl}


\bibitem[Chen et~al\mbox{.}(2021)]%
        {HumanEval}
\bibfield{author}{\bibinfo{person}{Mark Chen}, \bibinfo{person}{Jerry Tworek}, \bibinfo{person}{Heewoo Jun}, \bibinfo{person}{Qiming Yuan}, \bibinfo{person}{Henrique~Ponde de Oliveira~Pinto}, \bibinfo{person}{Jared Kaplan}, \bibinfo{person}{Harri Edwards}, \bibinfo{person}{Yuri Burda}, \bibinfo{person}{Nicholas Joseph}, \bibinfo{person}{Greg Brockman}, \bibinfo{person}{Alex Ray}, \bibinfo{person}{Raul Puri}, \bibinfo{person}{Gretchen Krueger}, \bibinfo{person}{Michael Petrov}, \bibinfo{person}{Heidy Khlaaf}, \bibinfo{person}{Girish Sastry}, \bibinfo{person}{Pamela Mishkin}, \bibinfo{person}{Brooke Chan}, \bibinfo{person}{Scott Gray}, \bibinfo{person}{Nick Ryder}, \bibinfo{person}{Mikhail Pavlov}, \bibinfo{person}{Alethea Power}, \bibinfo{person}{Lukasz Kaiser}, \bibinfo{person}{Mohammad Bavarian}, \bibinfo{person}{Clemens Winter}, \bibinfo{person}{Philippe Tillet}, \bibinfo{person}{Felipe~Petroski Such}, \bibinfo{person}{Dave Cummings}, \bibinfo{person}{Matthias Plappert}, \bibinfo{person}{Fotios Chantzis},
  \bibinfo{person}{Elizabeth Barnes}, \bibinfo{person}{Ariel Herbert-Voss}, \bibinfo{person}{William~Hebgen Guss}, \bibinfo{person}{Alex Nichol}, \bibinfo{person}{Alex Paino}, \bibinfo{person}{Nikolas Tezak}, \bibinfo{person}{Jie Tang}, \bibinfo{person}{Igor Babuschkin}, \bibinfo{person}{Suchir Balaji}, \bibinfo{person}{Shantanu Jain}, \bibinfo{person}{William Saunders}, \bibinfo{person}{Christopher Hesse}, \bibinfo{person}{Andrew~N. Carr}, \bibinfo{person}{Jan Leike}, \bibinfo{person}{Josh Achiam}, \bibinfo{person}{Vedant Misra}, \bibinfo{person}{Evan Morikawa}, \bibinfo{person}{Alec Radford}, \bibinfo{person}{Matthew Knight}, \bibinfo{person}{Miles Brundage}, \bibinfo{person}{Mira Murati}, \bibinfo{person}{Katie Mayer}, \bibinfo{person}{Peter Welinder}, \bibinfo{person}{Bob McGrew}, \bibinfo{person}{Dario Amodei}, \bibinfo{person}{Sam McCandlish}, \bibinfo{person}{Ilya Sutskever}, {and} \bibinfo{person}{Wojciech Zaremba}.} \bibinfo{year}{2021}\natexlab{}.
\newblock \showarticletitle{Evaluating Large Language Models Trained on Code}.
\newblock  (\bibinfo{year}{2021}).
\newblock
\showeprint[arxiv]{2107.03374}~[cs.LG]


\bibitem[Cobbe et~al\mbox{.}(2021)]%
        {gsm8k}
\bibfield{author}{\bibinfo{person}{Karl Cobbe}, \bibinfo{person}{Vineet Kosaraju}, \bibinfo{person}{Mohammad Bavarian}, \bibinfo{person}{Mark Chen}, \bibinfo{person}{Heewoo Jun}, \bibinfo{person}{Lukasz Kaiser}, \bibinfo{person}{Matthias Plappert}, \bibinfo{person}{Jerry Tworek}, \bibinfo{person}{Jacob Hilton}, \bibinfo{person}{Reiichiro Nakano}, \bibinfo{person}{Christopher Hesse}, {and} \bibinfo{person}{John Schulman}.} \bibinfo{year}{2021}\natexlab{}.
\newblock \showarticletitle{Training Verifiers to Solve Math Word Problems}.
\newblock \bibinfo{journal}{\emph{arXiv preprint arXiv:2110.14168}} (\bibinfo{year}{2021}).
\newblock


\bibitem[Cui et~al\mbox{.}(2023)]%
        {UltraFeedback}
\bibfield{author}{\bibinfo{person}{Ganqu Cui}, \bibinfo{person}{Liangyuan Yuan}, \bibinfo{person}{Ning Ding}, \bibinfo{person}{Zhiwei Yao}, \bibinfo{person}{Wei Ye}, \bibinfo{person}{Yujia Wang}, \bibinfo{person}{Yue Zhang}, \bibinfo{person}{Jing Xu}, \bibinfo{person}{Han Zhang}, \bibinfo{person}{Zini Chen}, {et~al\mbox{.}}} \bibinfo{year}{2023}\natexlab{}.
\newblock \showarticletitle{UltraFeedback: Boosting Language Models with High-quality Feedback}.
\newblock \bibinfo{journal}{\emph{arXiv preprint arXiv:2310.01377}} (\bibinfo{year}{2023}).
\newblock


\bibitem[Ding(2026)]%
        {HDPO}
\bibfield{author}{\bibinfo{person}{Ken Ding}.} \bibinfo{year}{2026}\natexlab{}.
\newblock \showarticletitle{HDPO: Hybrid Distillation Policy Optimization via Privileged Self-Distillation}.
\newblock \bibinfo{journal}{\emph{arXiv preprint arXiv:2603.23871}} (\bibinfo{year}{2026}).
\newblock


\bibitem[Gao et~al\mbox{.}(2025)]%
        {sapo}
\bibfield{author}{\bibinfo{person}{Chang Gao}, \bibinfo{person}{Chujie Zheng}, \bibinfo{person}{Xiong-Hui Chen}, \bibinfo{person}{Kai Dang}, \bibinfo{person}{Shixuan Liu}, \bibinfo{person}{Bowen Yu}, \bibinfo{person}{An Yang}, \bibinfo{person}{Shuai Bai}, \bibinfo{person}{Jingren Zhou}, {and} \bibinfo{person}{Junyang Lin}.} \bibinfo{year}{2025}\natexlab{}.
\newblock \showarticletitle{Soft adaptive policy optimization}.
\newblock \bibinfo{journal}{\emph{arXiv preprint arXiv:2511.20347}} (\bibinfo{year}{2025}).
\newblock


\bibitem[Guha et~al\mbox{.}(2025)]%
        {OpenThoughts}
\bibfield{author}{\bibinfo{person}{Etash Guha}, \bibinfo{person}{Ryan Marten}, \bibinfo{person}{Sedrick Keh}, \bibinfo{person}{Negin Raoof}, \bibinfo{person}{Georgios Smyrnis}, \bibinfo{person}{Hritik Bansal}, \bibinfo{person}{Marianna Nezhurina}, \bibinfo{person}{Jean Mercat}, \bibinfo{person}{Trung Vu}, \bibinfo{person}{Zayne Sprague}, \bibinfo{person}{Ashima Suvarna}, \bibinfo{person}{Benjamin Feuer}, \bibinfo{person}{Liangyu Chen}, \bibinfo{person}{Zaid Khan}, \bibinfo{person}{Eric Frankel}, \bibinfo{person}{Sachin Grover}, \bibinfo{person}{Caroline Choi}, \bibinfo{person}{Niklas Muennighoff}, \bibinfo{person}{Shiye Su}, \bibinfo{person}{Wanjia Zhao}, \bibinfo{person}{John Yang}, \bibinfo{person}{Shreyas Pimpalgaonkar}, \bibinfo{person}{Kartik Sharma}, \bibinfo{person}{Charlie Cheng-Jie Ji}, \bibinfo{person}{Yichuan Deng}, \bibinfo{person}{Sarah Pratt}, \bibinfo{person}{Vivek Ramanujan}, \bibinfo{person}{Jon Saad-Falcon}, \bibinfo{person}{Jeffrey Li}, \bibinfo{person}{Achal Dave}, \bibinfo{person}{Alon
  Albalak}, \bibinfo{person}{Kushal Arora}, \bibinfo{person}{Blake Wulfe}, \bibinfo{person}{Chinmay Hegde}, \bibinfo{person}{Greg Durrett}, \bibinfo{person}{Sewoong Oh}, \bibinfo{person}{Mohit Bansal}, \bibinfo{person}{Saadia Gabriel}, \bibinfo{person}{Aditya Grover}, \bibinfo{person}{Kai-Wei Chang}, \bibinfo{person}{Vaishaal Shankar}, \bibinfo{person}{Aaron Gokaslan}, \bibinfo{person}{Mike~A. Merrill}, \bibinfo{person}{Tatsunori Hashimoto}, \bibinfo{person}{Yejin Choi}, \bibinfo{person}{Jenia Jitsev}, \bibinfo{person}{Reinhard Heckel}, \bibinfo{person}{Maheswaran Sathiamoorthy}, \bibinfo{person}{Alexandros~G. Dimakis}, {and} \bibinfo{person}{Ludwig Schmidt}.} \bibinfo{year}{2025}\natexlab{}.
\newblock \bibinfo{title}{OpenThoughts: Data Recipes for Reasoning Models}.
\newblock
\showeprint[arxiv]{2506.04178}~[cs.LG]
\urldef\tempurl%
\url{https://arxiv.org/abs/2506.04178}
\showURL{%
\tempurl}


\bibitem[Han et~al\mbox{.}(2026)]%
        {Adaptive}
\bibfield{author}{\bibinfo{person}{Zihao Han}, \bibinfo{person}{Tiangang Zhang}, \bibinfo{person}{Huaibin Wang}, {and} \bibinfo{person}{Yilun Sun}.} \bibinfo{year}{2026}\natexlab{}.
\newblock \bibinfo{title}{Adaptive Teacher Exposure for Self-Distillation in LLM Reasoning}.
\newblock
\showeprint[arxiv]{2605.11458}~[cs.AI]
\urldef\tempurl%
\url{https://arxiv.org/abs/2605.11458}
\showURL{%
\tempurl}


\bibitem[He et~al\mbox{.}(2026)]%
        {Self-Distillation-Zero}
\bibfield{author}{\bibinfo{person}{Yinghui He}, \bibinfo{person}{Simran Kaur}, \bibinfo{person}{Adithya Bhaskar}, \bibinfo{person}{Yongjin Yang}, \bibinfo{person}{Jiarui Liu}, \bibinfo{person}{Narutatsu Ri}, \bibinfo{person}{Liam Fowl}, \bibinfo{person}{Abhishek Panigrahi}, \bibinfo{person}{Danqi Chen}, {and} \bibinfo{person}{Sanjeev Arora}.} \bibinfo{year}{2026}\natexlab{}.
\newblock \showarticletitle{Self-Distillation Zero: Self-Revision Turns Binary Rewards into Dense Supervision}.
\newblock \bibinfo{journal}{\emph{arXiv preprint arXiv:2604.12002}} (\bibinfo{year}{2026}).
\newblock
\urldef\tempurl%
\url{https://arxiv.org/abs/2604.12002}
\showURL{%
\tempurl}


\bibitem[Hendrycks et~al\mbox{.}(2021)]%
        {math}
\bibfield{author}{\bibinfo{person}{Dan Hendrycks}, \bibinfo{person}{Collin Burns}, \bibinfo{person}{Saurav Kadavath}, \bibinfo{person}{Akul Arora}, \bibinfo{person}{Steven Basart}, \bibinfo{person}{Eric Tang}, \bibinfo{person}{Dawn Song}, {and} \bibinfo{person}{Jacob Steinhardt}.} \bibinfo{year}{2021}\natexlab{}.
\newblock \showarticletitle{Measuring Mathematical Problem Solving With the {MATH} Dataset}.
\newblock \bibinfo{journal}{\emph{Advances in Neural Information Processing Systems}}  \bibinfo{volume}{34} (\bibinfo{year}{2021}), \bibinfo{pages}{25774--25786}.
\newblock


\bibitem[HuggingFaceH4(2024)]%
        {aime}
\bibfield{author}{\bibinfo{person}{HuggingFaceH4}.} \bibinfo{year}{2024}\natexlab{}.
\newblock \bibinfo{title}{AIME 2024 Dataset}.
\newblock \bibinfo{howpublished}{\url{https://huggingface.co/datasets/HuggingFaceH4/aime_2024}}.
\newblock


\bibitem[Hübotter et~al\mbox{.}(2026)]%
        {Reinforcement}
\bibfield{author}{\bibinfo{person}{Jonas Hübotter}, \bibinfo{person}{Frederike Lübeck}, \bibinfo{person}{Lejs Behric}, \bibinfo{person}{Anton Baumann}, \bibinfo{person}{Marco Bagatella}, \bibinfo{person}{Daniel Marta}, \bibinfo{person}{Ido Hakimi}, \bibinfo{person}{Idan Shenfeld}, \bibinfo{person}{Thomas~Kleine Buening}, \bibinfo{person}{Carlos Guestrin}, {and} \bibinfo{person}{Andreas Krause}.} \bibinfo{year}{2026}\natexlab{}.
\newblock \bibinfo{title}{Reinforcement Learning via Self-Distillation}.
\newblock
\showeprint[arxiv]{2601.20802}~[cs.LG]
\urldef\tempurl%
\url{https://arxiv.org/abs/2601.20802}
\showURL{%
\tempurl}


\bibitem[Jain et~al\mbox{.}(2024)]%
        {livecodebench}
\bibfield{author}{\bibinfo{person}{Naman Jain}, \bibinfo{person}{King Han}, \bibinfo{person}{Alex Gu}, \bibinfo{person}{Wen-Ding Li}, \bibinfo{person}{Fanjia Yan}, \bibinfo{person}{Tianjun Zhang}, \bibinfo{person}{Sida Wang}, \bibinfo{person}{Armando Solar-Lezama}, \bibinfo{person}{Koushik Sen}, {and} \bibinfo{person}{Ion Stoica}.} \bibinfo{year}{2024}\natexlab{}.
\newblock \showarticletitle{LiveCodeBench: Holistic and Contamination Free Evaluation of Large Language Models for Code}.
\newblock \bibinfo{journal}{\emph{arXiv preprint}} (\bibinfo{year}{2024}).
\newblock


\bibitem[Jeong(2026)]%
        {Healthcare}
\bibfield{author}{\bibinfo{person}{Minbyul Jeong}.} \bibinfo{year}{2026}\natexlab{}.
\newblock \bibinfo{title}{Healthcare AI GYM for Medical Agents}.
\newblock
\showeprint[arxiv]{2605.02943}~[cs.LG]
\urldef\tempurl%
\url{https://arxiv.org/abs/2605.02943}
\showURL{%
\tempurl}


\bibitem[Ji et~al\mbox{.}(2023)]%
        {BeaverTails}
\bibfield{author}{\bibinfo{person}{Jiaming Ji}, \bibinfo{person}{Meng Liu}, \bibinfo{person}{Juntao Dai}, \bibinfo{person}{Xuehai Pan}, \bibinfo{person}{Ce Zhang}, \bibinfo{person}{Chi Bian}, \bibinfo{person}{Botao Chen}, \bibinfo{person}{Rui Sun}, \bibinfo{person}{Yashi Wang}, {and} \bibinfo{person}{Yaodong Yang}.} \bibinfo{year}{2023}\natexlab{}.
\newblock \showarticletitle{BeaverTails: Towards Improved Safety Alignment of LLM via a Human-Preference Dataset}.
\newblock \bibinfo{journal}{\emph{Advances in Neural Information Processing Systems}}  \bibinfo{volume}{36} (\bibinfo{year}{2023}), \bibinfo{pages}{24621--24658}.
\newblock


\bibitem[Jiang et~al\mbox{.}(2026)]%
        {D-OPSD}
\bibfield{author}{\bibinfo{person}{Dengyang Jiang}, \bibinfo{person}{Xin Jin}, \bibinfo{person}{Dongyang Liu}, \bibinfo{person}{Zanyi Wang}, \bibinfo{person}{Mingzhe Zheng}, \bibinfo{person}{Ruoyi Du}, \bibinfo{person}{Xiangpeng Yang}, \bibinfo{person}{Qilong Wu}, \bibinfo{person}{Zhen Li}, \bibinfo{person}{Peng Gao}, \bibinfo{person}{Harry Yang}, {and} \bibinfo{person}{Steven Hoi}.} \bibinfo{year}{2026}\natexlab{}.
\newblock \bibinfo{title}{D-OPSD: On-Policy Self-Distillation for Continuously Tuning Step-Distilled Diffusion Models}.
\newblock
\showeprint[arxiv]{2605.05204}~[cs.CV]
\urldef\tempurl%
\url{https://arxiv.org/abs/2605.05204}
\showURL{%
\tempurl}


\bibitem[Jiang et~al\mbox{.}(2025)]%
        {sft4}
\bibfield{author}{\bibinfo{person}{Tingyu Jiang}, \bibinfo{person}{Shen Li}, \bibinfo{person}{Yiyao Song}, \bibinfo{person}{Lan Zhang}, \bibinfo{person}{Hualei Zhu}, \bibinfo{person}{Yuan Zhao}, \bibinfo{person}{Xiaohang Xu}, \bibinfo{person}{Kenjiro Taura}, {and} \bibinfo{person}{Hao~Henry Wang}.} \bibinfo{year}{2025}\natexlab{}.
\newblock \bibinfo{title}{Importance-Aware Data Selection for Efficient LLM Instruction Tuning}.
\newblock
\showeprint[arxiv]{2511.07074}~[cs.CL]
\urldef\tempurl%
\url{https://arxiv.org/abs/2511.07074}
\showURL{%
\tempurl}


\bibitem[Jin et~al\mbox{.}(2026)]%
        {UniSD}
\bibfield{author}{\bibinfo{person}{Yiqiao Jin}, \bibinfo{person}{Yiyang Wang}, \bibinfo{person}{Lucheng Fu}, \bibinfo{person}{Yijia Xiao}, \bibinfo{person}{Yinyi Luo}, \bibinfo{person}{Haoxin Liu}, \bibinfo{person}{B.~Aditya Prakash}, \bibinfo{person}{Josiah Hester}, \bibinfo{person}{Jindong Wang}, {and} \bibinfo{person}{Srijan Kumar}.} \bibinfo{year}{2026}\natexlab{}.
\newblock \bibinfo{title}{UniSD: Towards a Unified Self-Distillation Framework for Large Language Models}.
\newblock
\showeprint[arxiv]{2605.06597}~[cs.CL]
\urldef\tempurl%
\url{https://arxiv.org/abs/2605.06597}
\showURL{%
\tempurl}


\bibitem[Ke et~al\mbox{.}(2026)]%
        {Respecting}
\bibfield{author}{\bibinfo{person}{Junlong Ke}, \bibinfo{person}{Zichen Wen}, \bibinfo{person}{Weijia Li}, \bibinfo{person}{Conghui He}, {and} \bibinfo{person}{Linfeng Zhang}.} \bibinfo{year}{2026}\natexlab{}.
\newblock \bibinfo{title}{Respecting Self-Uncertainty in On-Policy Self-Distillation for Efficient LLM Reasoning}.
\newblock
\showeprint[arxiv]{2605.13255}~[cs.AI]
\urldef\tempurl%
\url{https://arxiv.org/abs/2605.13255}
\showURL{%
\tempurl}


\bibitem[Kim et~al\mbox{.}(2026a)]%
        {Rebellious}
\bibfield{author}{\bibinfo{person}{Jeonghye Kim}, \bibinfo{person}{Jiwon Jeon}, \bibinfo{person}{Dongsheng Li}, {and} \bibinfo{person}{Yuqing Yang}.} \bibinfo{year}{2026}\natexlab{a}.
\newblock \bibinfo{title}{Rebellious Student: Reversing Teacher Signals for Reasoning Exploration with Self-Distilled RLVR}.
\newblock
\showeprint[arxiv]{2605.10781}~[cs.LG]
\urldef\tempurl%
\url{https://arxiv.org/abs/2605.10781}
\showURL{%
\tempurl}


\bibitem[Kim et~al\mbox{.}(2026b)]%
        {why}
\bibfield{author}{\bibinfo{person}{Jeonghye Kim}, \bibinfo{person}{Xufang Luo}, \bibinfo{person}{Minbeom Kim}, \bibinfo{person}{Sangmook Lee}, \bibinfo{person}{Dohyung Kim}, \bibinfo{person}{Jiwon Jeon}, \bibinfo{person}{Dongsheng Li}, {and} \bibinfo{person}{Yuqing Yang}.} \bibinfo{year}{2026}\natexlab{b}.
\newblock \bibinfo{title}{Why Does Self-Distillation (Sometimes) Degrade the Reasoning Capability of LLMs?}
\newblock
\showeprint[arxiv]{2603.24472}~[cs.CL]
\urldef\tempurl%
\url{https://arxiv.org/abs/2603.24472}
\showURL{%
\tempurl}


\bibitem[Kirchenbauer et~al\mbox{.}(2026)]%
        {Multi-Token}
\bibfield{author}{\bibinfo{person}{John Kirchenbauer}, \bibinfo{person}{Abhimanyu Hans}, \bibinfo{person}{Brian Bartoldson}, \bibinfo{person}{Micah Goldblum}, \bibinfo{person}{Ashwinee Panda}, {and} \bibinfo{person}{Tom Goldstein}.} \bibinfo{year}{2026}\natexlab{}.
\newblock \bibinfo{title}{Multi-Token Prediction via Self-Distillation}.
\newblock
\showeprint[arxiv]{2602.06019}~[cs.CL]
\urldef\tempurl%
\url{https://arxiv.org/abs/2602.06019}
\showURL{%
\tempurl}


\bibitem[Li et~al\mbox{.}(2026b)]%
        {Unifying}
\bibfield{author}{\bibinfo{person}{Gengsheng Li}, \bibinfo{person}{Tianyu Yang}, \bibinfo{person}{Junfeng Fang}, \bibinfo{person}{Mingyang Song}, \bibinfo{person}{Mao Zheng}, \bibinfo{person}{Haiyun Guo}, \bibinfo{person}{Dan Zhang}, \bibinfo{person}{Jinqiao Wang}, {and} \bibinfo{person}{Tat-Seng Chua}.} \bibinfo{year}{2026}\natexlab{b}.
\newblock \bibinfo{title}{Unifying Group-Relative and Self-Distillation Policy Optimization via Sample Routing}.
\newblock
\showeprint[arxiv]{2604.02288}~[cs.LG]
\urldef\tempurl%
\url{https://arxiv.org/abs/2604.02288}
\showURL{%
\tempurl}


\bibitem[Li et~al\mbox{.}(2026a)]%
        {GEAR}
\bibfield{author}{\bibinfo{person}{Sijia Li}, \bibinfo{person}{Yuchen Huang}, \bibinfo{person}{Zifan Liu}, \bibinfo{person}{Yanping Li}, \bibinfo{person}{Jingjing Fu}, \bibinfo{person}{Li Zhao}, \bibinfo{person}{Jiang Bian}, \bibinfo{person}{Ling Zhang}, \bibinfo{person}{Jun Zhang}, {and} \bibinfo{person}{Rui Wang}.} \bibinfo{year}{2026}\natexlab{a}.
\newblock \bibinfo{title}{GEAR: Granularity-Adaptive Advantage Reweighting for LLM Agents via Self-Distillation}.
\newblock
\showeprint[arxiv]{2605.11853}~[cs.LG]
\urldef\tempurl%
\url{https://arxiv.org/abs/2605.11853}
\showURL{%
\tempurl}


\bibitem[Li et~al\mbox{.}(2026c)]%
        {Survey2}
\bibfield{author}{\bibinfo{person}{Yaxuan Li}, \bibinfo{person}{Yuxin Zuo}, \bibinfo{person}{Bingxiang He}, \bibinfo{person}{Jinqian Zhang}, \bibinfo{person}{Chaojun Xiao}, \bibinfo{person}{Cheng Qian}, \bibinfo{person}{Tianyu Yu}, \bibinfo{person}{Huan ang Gao}, \bibinfo{person}{Wenkai Yang}, \bibinfo{person}{Zhiyuan Liu}, {and} \bibinfo{person}{Ning Ding}.} \bibinfo{year}{2026}\natexlab{c}.
\newblock \bibinfo{title}{Rethinking On-Policy Distillation of Large Language Models: Phenomenology, Mechanism, and Recipe}.
\newblock
\showeprint[arxiv]{2604.13016}~[cs.LG]
\urldef\tempurl%
\url{https://arxiv.org/abs/2604.13016}
\showURL{%
\tempurl}


\bibitem[Liao et~al\mbox{.}(2026)]%
        {IRIS}
\bibfield{author}{\bibinfo{person}{Wenjie Liao}, \bibinfo{person}{Like Wu}, \bibinfo{person}{Liangjie Zhao}, \bibinfo{person}{Shihui Xu}, {and} \bibinfo{person}{Shigeru Fujimura}.} \bibinfo{year}{2026}\natexlab{}.
\newblock \bibinfo{title}{IRIS: Interpolative R\'enyi Iterative Self-play for Large Language Model Fine-Tuning}.
\newblock
\showeprint[arxiv]{2604.20933}~[cs.LG]
\urldef\tempurl%
\url{https://arxiv.org/abs/2604.20933}
\showURL{%
\tempurl}


\bibitem[Lin et~al\mbox{.}(2026)]%
        {VISD}
\bibfield{author}{\bibinfo{person}{Hao Lin}, \bibinfo{person}{Kunyang Lv}, \bibinfo{person}{Xu Jiang}, \bibinfo{person}{Jingqi Tian}, \bibinfo{person}{Zhongjing Du}, \bibinfo{person}{Jiayu Ding}, \bibinfo{person}{Qiaoman Zhang}, {and} \bibinfo{person}{Hongbo Jin}.} \bibinfo{year}{2026}\natexlab{}.
\newblock \bibinfo{title}{VISD: Enhancing Video Reasoning via Structured Self-Distillation}.
\newblock
\showeprint[arxiv]{2605.06094}~[cs.CV]
\urldef\tempurl%
\url{https://arxiv.org/abs/2605.06094}
\showURL{%
\tempurl}


\bibitem[Lin et~al\mbox{.}(2014)]%
        {coco}
\bibfield{author}{\bibinfo{person}{Tsung-Yi Lin}, \bibinfo{person}{Michael Maire}, \bibinfo{person}{Serge Belongie}, \bibinfo{person}{James Hays}, \bibinfo{person}{Pietro Perona}, \bibinfo{person}{Deva Ramanan}, \bibinfo{person}{Piotr Doll{\'a}r}, {and} \bibinfo{person}{C~Lawrence Zitnick}.} \bibinfo{year}{2014}\natexlab{}.
\newblock \showarticletitle{Microsoft coco: Common objects in context}.
\newblock \bibinfo{journal}{\emph{arXiv preprint arXiv:1405.0312}} (\bibinfo{year}{2014}).
\newblock


\bibitem[Liu et~al\mbox{.}(2026)]%
        {Crosslingual}
\bibfield{author}{\bibinfo{person}{Yihong Liu}, \bibinfo{person}{Raoyuan Zhao}, \bibinfo{person}{Michael~A. Hedderich}, {and} \bibinfo{person}{Hinrich Schütze}.} \bibinfo{year}{2026}\natexlab{}.
\newblock \bibinfo{title}{Crosslingual On-Policy Self-Distillation for Multilingual Reasoning}.
\newblock
\showeprint[arxiv]{2605.09548}~[cs.CL]
\urldef\tempurl%
\url{https://arxiv.org/abs/2605.09548}
\showURL{%
\tempurl}


\bibitem[Lu and Lab(2025)]%
        {opd2}
\bibfield{author}{\bibinfo{person}{Kevin Lu} {and} \bibinfo{person}{Thinking~Machines Lab}.} \bibinfo{year}{2025}\natexlab{}.
\newblock \showarticletitle{On-Policy Distillation}.
\newblock \bibinfo{journal}{\emph{Thinking Machines Lab: Connectionism}} (\bibinfo{year}{2025}).
\newblock
\href{https://doi.org/10.64434/tml.20251026}{doi:\nolinkurl{10.64434/tml.20251026}}
\newblock
\shownote{https://thinkingmachines.ai/blog/on-policy-distillation}.


\bibitem[Ouyang et~al\mbox{.}(2022)]%
        {sft1}
\bibfield{author}{\bibinfo{person}{Long Ouyang}, \bibinfo{person}{Jeff Wu}, \bibinfo{person}{Xu Jiang}, \bibinfo{person}{Diogo Almeida}, \bibinfo{person}{Carroll~L. Wainwright}, \bibinfo{person}{Pamela Mishkin}, \bibinfo{person}{Chong Zhang}, \bibinfo{person}{Sandhini Agarwal}, \bibinfo{person}{Katarina Slama}, \bibinfo{person}{Alex Ray}, \bibinfo{person}{John Schulman}, \bibinfo{person}{Jacob Hilton}, \bibinfo{person}{Fraser Kelton}, \bibinfo{person}{Luke Miller}, \bibinfo{person}{Maddie Simens}, \bibinfo{person}{Amanda Askell}, \bibinfo{person}{Peter Welinder}, \bibinfo{person}{Paul Christiano}, \bibinfo{person}{Jan Leike}, {and} \bibinfo{person}{Ryan Lowe}.} \bibinfo{year}{2022}\natexlab{}.
\newblock \bibinfo{title}{Training language models to follow instructions with human feedback}.
\newblock
\showeprint[arxiv]{2203.02155}~[cs.CL]
\urldef\tempurl%
\url{https://arxiv.org/abs/2203.02155}
\showURL{%
\tempurl}


\bibitem[Penaloza et~al\mbox{.}(2026)]%
        {Privileged}
\bibfield{author}{\bibinfo{person}{Emiliano Penaloza}, \bibinfo{person}{Dheeraj Vattikonda}, \bibinfo{person}{Nicolas Gontier}, \bibinfo{person}{Alexandre Lacoste}, \bibinfo{person}{Laurent Charlin}, {and} \bibinfo{person}{Massimo Caccia}.} \bibinfo{year}{2026}\natexlab{}.
\newblock \showarticletitle{Privileged Information Distillation for Language Models}.
\newblock \bibinfo{journal}{\emph{arXiv preprint arXiv:2602.04942}} (\bibinfo{year}{2026}).
\newblock


\bibitem[Qin et~al\mbox{.}(2026)]%
        {Multilingual}
\bibfield{author}{\bibinfo{person}{Ruiyang Qin}, \bibinfo{person}{Qingzhuo Wang}, \bibinfo{person}{Dongrui Liu}, \bibinfo{person}{Qiang Li}, \bibinfo{person}{Zhihua Wei}, {and} \bibinfo{person}{Wen Shen}.} \bibinfo{year}{2026}\natexlab{}.
\newblock \bibinfo{title}{Multilingual Safety Alignment via Self-Distillation}.
\newblock
\showeprint[arxiv]{2605.02971}~[cs.LG]
\urldef\tempurl%
\url{https://arxiv.org/abs/2605.02971}
\showURL{%
\tempurl}


\bibitem[Sang et~al\mbox{.}(2026)]%
        {On-Policy-Self}
\bibfield{author}{\bibinfo{person}{Hejian Sang}, \bibinfo{person}{Yuanda Xu}, \bibinfo{person}{Zhengze Zhou}, \bibinfo{person}{Ran He}, \bibinfo{person}{Zhipeng Wang}, {and} \bibinfo{person}{Jiachen Sun}.} \bibinfo{year}{2026}\natexlab{}.
\newblock \showarticletitle{On-Policy Self-Distillation for Reasoning Compression}.
\newblock \bibinfo{journal}{\emph{arXiv preprint arXiv:2603.05433}} (\bibinfo{year}{2026}).
\newblock


\bibitem[Shao et~al\mbox{.}(024b)]%
        {shao2022deepseek}
\bibfield{author}{\bibinfo{person}{Zhihong Shao}, \bibinfo{person}{Peiyi Wang}, \bibinfo{person}{Qihao Zhu}, \bibinfo{person}{Runxin Xu}, \bibinfo{person}{Junxiao Song}, \bibinfo{person}{Xiao Bi}, \bibinfo{person}{Mingchuan Zhang}, \bibinfo{person}{Y.~K. Li}, \bibinfo{person}{Y. Wu}, {and} \bibinfo{person}{Daya Guo}.} \bibinfo{year}{2024b}\natexlab{}.
\newblock \showarticletitle{Deepseekmath: Pushing the limits of mathematical reasoning in open language models}.
\newblock \bibinfo{journal}{\emph{arXiv preprint arXiv:2402.03300}} (\bibinfo{year}{2024b}).
\newblock


\bibitem[Shen et~al\mbox{.}(2026a)]%
        {Anti-Self-Distillation}
\bibfield{author}{\bibinfo{person}{Guobin Shen}, \bibinfo{person}{Xiang Cheng}, \bibinfo{person}{Chenxiao Zhao}, \bibinfo{person}{Lei Huang}, \bibinfo{person}{Jindong Li}, \bibinfo{person}{Dongcheng Zhao}, {and} \bibinfo{person}{Xing Yu}.} \bibinfo{year}{2026}\natexlab{a}.
\newblock \bibinfo{title}{Anti-Self-Distillation for Reasoning RL via Pointwise Mutual Information}.
\newblock
\showeprint[arxiv]{2605.11609}~[cs.LG]
\urldef\tempurl%
\url{https://arxiv.org/abs/2605.11609}
\showURL{%
\tempurl}


\bibitem[Shen et~al\mbox{.}(2026b)]%
        {From}
\bibfield{author}{\bibinfo{person}{Guobin Shen}, \bibinfo{person}{Lei Huang}, \bibinfo{person}{Xiang Cheng}, \bibinfo{person}{Chenxiao Zhao}, \bibinfo{person}{Jindong Li}, \bibinfo{person}{Dongcheng Zhao}, {and} \bibinfo{person}{Xing Yu}.} \bibinfo{year}{2026}\natexlab{b}.
\newblock \bibinfo{title}{From Generic Correlation to Input-Specific Credit in On-Policy Self Distillation}.
\newblock
\showeprint[arxiv]{2605.11613}~[cs.LG]
\urldef\tempurl%
\url{https://arxiv.org/abs/2605.11613}
\showURL{%
\tempurl}


\bibitem[Shenfeld et~al\mbox{.}(2026)]%
        {Self-Distillation}
\bibfield{author}{\bibinfo{person}{Idan Shenfeld}, \bibinfo{person}{Mehul Damani}, \bibinfo{person}{Jonas Hübotter}, {and} \bibinfo{person}{Pulkit Agrawal}.} \bibinfo{year}{2026}\natexlab{}.
\newblock \bibinfo{title}{Self-Distillation Enables Continual Learning}.
\newblock
\showeprint[arxiv]{2601.19897}~[cs.LG]
\urldef\tempurl%
\url{https://arxiv.org/abs/2601.19897}
\showURL{%
\tempurl}


\bibitem[Shridhar et~al\mbox{.}(2021)]%
        {ALFWorld}
\bibfield{author}{\bibinfo{person}{Mohit Shridhar}, \bibinfo{person}{Xingdi Yuan}, \bibinfo{person}{Marc-Alexandre Cote}, \bibinfo{person}{Yonatan Bisk}, \bibinfo{person}{Adam Trischler}, {and} \bibinfo{person}{Matthew Hausknecht}.} \bibinfo{year}{2021}\natexlab{}.
\newblock \showarticletitle{ALFWorld: Aligning Text and Embodied Environments for Interactive Learning}.
\newblock \bibinfo{journal}{\emph{arXiv preprint arXiv:2010.03768}} (\bibinfo{year}{2021}).
\newblock


\bibitem[Song and Zheng(2026)]%
        {Survey}
\bibfield{author}{\bibinfo{person}{Mingyang Song} {and} \bibinfo{person}{Mao Zheng}.} \bibinfo{year}{2026}\natexlab{}.
\newblock \bibinfo{title}{A Survey of On-Policy Distillation for Large Language Models}.
\newblock
\showeprint[arxiv]{2604.00626}~[cs.LG]
\urldef\tempurl%
\url{https://arxiv.org/abs/2604.00626}
\showURL{%
\tempurl}


\bibitem[Stein et~al\mbox{.}(2026)]%
        {GATES}
\bibfield{author}{\bibinfo{person}{Alex Stein}, \bibinfo{person}{Furong Huang}, {and} \bibinfo{person}{Tom Goldstein}.} \bibinfo{year}{2026}\natexlab{}.
\newblock \showarticletitle{GATES: Self-Distillation under Privileged Context with Consensus Gating}.
\newblock \bibinfo{journal}{\emph{arXiv preprint arXiv:2602.20574}} (\bibinfo{year}{2026}).
\newblock


\bibitem[Tan and Hong(2026)]%
        {PAINT}
\bibfield{author}{\bibinfo{person}{Zhiquan Tan} {and} \bibinfo{person}{Yinrong Hong}.} \bibinfo{year}{2026}\natexlab{}.
\newblock \bibinfo{title}{PAINT: Partial-Solution Adaptive Interpolated Training for Self-Distilled Reasoners}.
\newblock
\showeprint[arxiv]{2604.26573}~[cs.LG]
\urldef\tempurl%
\url{https://arxiv.org/abs/2604.26573}
\showURL{%
\tempurl}


\bibitem[Wang et~al\mbox{.}(2026)]%
        {Skill-SD}
\bibfield{author}{\bibinfo{person}{Hao Wang}, \bibinfo{person}{Guozhi Wang}, \bibinfo{person}{Han Xiao}, \bibinfo{person}{Yufeng Zhou}, \bibinfo{person}{Yue Pan}, \bibinfo{person}{Jichao Wang}, \bibinfo{person}{Ke Xu}, \bibinfo{person}{Yafei Wen}, \bibinfo{person}{Xiaohu Ruan}, \bibinfo{person}{Xiaoxin Chen}, {and} \bibinfo{person}{Honggang Qi}.} \bibinfo{year}{2026}\natexlab{}.
\newblock \bibinfo{title}{Skill-SD: Skill-Conditioned Self-Distillation for Multi-turn LLM Agents}.
\newblock
\showeprint[arxiv]{2604.10674}~[cs.LG]
\urldef\tempurl%
\url{https://arxiv.org/abs/2604.10674}
\showURL{%
\tempurl}


\bibitem[Xiao et~al\mbox{.}(2026)]%
        {sft3}
\bibfield{author}{\bibinfo{person}{Yuxin Xiao}, \bibinfo{person}{Shujian Zhang}, \bibinfo{person}{Wenxuan Zhou}, \bibinfo{person}{Marzyeh Ghassemi}, {and} \bibinfo{person}{Sanqiang Zhao}.} \bibinfo{year}{2026}\natexlab{}.
\newblock \bibinfo{title}{SFTMix: Elevating Language Model Instruction Tuning with Mixup Recipe}.
\newblock
\showeprint[arxiv]{2410.05248}~[cs.CL]
\urldef\tempurl%
\url{https://arxiv.org/abs/2410.05248}
\showURL{%
\tempurl}


\bibitem[Xu et~al\mbox{.}(2026)]%
        {PACED}
\bibfield{author}{\bibinfo{person}{Yuanda Xu}, \bibinfo{person}{Hejian Sang}, \bibinfo{person}{Zhengze Zhou}, \bibinfo{person}{Ran He}, {and} \bibinfo{person}{Zhipeng Wang}.} \bibinfo{year}{2026}\natexlab{}.
\newblock \bibinfo{title}{PACED: Distillation and On-Policy Self-Distillation at the Frontier of Student Competence}.
\newblock
\showeprint[arxiv]{2603.11178}~[cs.AI]
\urldef\tempurl%
\url{https://arxiv.org/abs/2603.11178}
\showURL{%
\tempurl}


\bibitem[Yang et~al\mbox{.}(2026a)]%
        {Self-Distilled-RLVR}
\bibfield{author}{\bibinfo{person}{Chenxu Yang}, \bibinfo{person}{Chuanyu Qin}, \bibinfo{person}{Qingyi Si}, \bibinfo{person}{Minghui Chen}, \bibinfo{person}{Naibin Gu}, \bibinfo{person}{Dingyu Yao}, \bibinfo{person}{Zheng Lin}, \bibinfo{person}{Weiping Wang}, \bibinfo{person}{Jiaqi Wang}, {and} \bibinfo{person}{Nan Duan}.} \bibinfo{year}{2026}\natexlab{a}.
\newblock \bibinfo{title}{Self-Distilled RLVR}.
\newblock
\showeprint[arxiv]{2604.03128}~[cs.LG]
\urldef\tempurl%
\url{https://arxiv.org/abs/2604.03128}
\showURL{%
\tempurl}


\bibitem[Yang et~al\mbox{.}(2026b)]%
        {OGLS-SD}
\bibfield{author}{\bibinfo{person}{Yuxiao Yang}, \bibinfo{person}{Xiaoyun Wang}, {and} \bibinfo{person}{Weitong Zhang}.} \bibinfo{year}{2026}\natexlab{b}.
\newblock \bibinfo{title}{OGLS-SD: On-Policy Self-Distillation with Outcome-Guided Logit Steering for LLM Reasoning}.
\newblock
\showeprint[arxiv]{2605.12400}~[cs.LG]
\urldef\tempurl%
\url{https://arxiv.org/abs/2605.12400}
\showURL{%
\tempurl}


\bibitem[Yao et~al\mbox{.}(2022)]%
        {Webshop}
\bibfield{author}{\bibinfo{person}{Shunyu Yao}, \bibinfo{person}{Howard Jiang}, \bibinfo{person}{Zilin Chen}, \bibinfo{person}{Chen Zhang}, \bibinfo{person}{Yichen Wang}, \bibinfo{person}{Ruocheng Chen}, \bibinfo{person}{Wenlong Gu}, \bibinfo{person}{Zipeng Zhang}, {and} \bibinfo{person}{Li Sha}.} \bibinfo{year}{2022}\natexlab{}.
\newblock \showarticletitle{Webshop: Towards Scalable Real-World Web Interaction with Grounded Language Agents}.
\newblock \bibinfo{journal}{\emph{arXiv preprint arXiv:2207.01206}} (\bibinfo{year}{2022}).
\newblock


\bibitem[Ye et~al\mbox{.}(2026a)]%
        {Online}
\bibfield{author}{\bibinfo{person}{Tianzhu Ye}, \bibinfo{person}{Li Dong}, \bibinfo{person}{Qingxiu Dong}, \bibinfo{person}{Xun Wu}, \bibinfo{person}{Shaohan Huang}, {and} \bibinfo{person}{Furu Wei}.} \bibinfo{year}{2026}\natexlab{a}.
\newblock \showarticletitle{Online Experiential Learning for Language Models}.
\newblock \bibinfo{journal}{\emph{arXiv preprint arXiv:2603.16856}} (\bibinfo{year}{2026}).
\newblock


\bibitem[Ye et~al\mbox{.}(2026b)]%
        {On-policy}
\bibfield{author}{\bibinfo{person}{Tianzhu Ye}, \bibinfo{person}{Li Dong}, \bibinfo{person}{Xun Wu}, \bibinfo{person}{Shaohan Huang}, {and} \bibinfo{person}{Furu Wei}.} \bibinfo{year}{2026}\natexlab{b}.
\newblock \showarticletitle{On-policy context distillation for language models}.
\newblock \bibinfo{journal}{\emph{arXiv preprint arXiv:2602.12275}} (\bibinfo{year}{2026}).
\newblock


\bibitem[Yu et~al\mbox{.}(025e)]%
        {yu2025dapo}
\bibfield{author}{\bibinfo{person}{Qiying Yu}, \bibinfo{person}{Zheng Zhang}, \bibinfo{person}{Ruofei Zhu}, \bibinfo{person}{Yufeng Yuan}, \bibinfo{person}{Xiaochen Zuo}, \bibinfo{person}{Yu Yu}, \bibinfo{person}{Weinan Dai}, \bibinfo{person}{TianTian Fan}, \bibinfo{person}{Gaohong Liu}, \bibinfo{person}{Lingjun Liu}, {and} \bibinfo{person}{et al.}} \bibinfo{year}{2025e}\natexlab{}.
\newblock \showarticletitle{Dapo: An open-source llm reinforcement learning system at scale}.
\newblock \bibinfo{journal}{\emph{arXiv preprint arXiv:2503.14476}} (\bibinfo{year}{2025e}).
\newblock


\bibitem[Yu et~al\mbox{.}(2026)]%
        {Preference}
\bibfield{author}{\bibinfo{person}{Xin Yu}, \bibinfo{person}{Liuchen Liao}, \bibinfo{person}{Yiwen Zhang}, \bibinfo{person}{Yingchen Yu}, \bibinfo{person}{Lingzhou Xue}, {and} \bibinfo{person}{Qinzhen Guo}.} \bibinfo{year}{2026}\natexlab{}.
\newblock \bibinfo{title}{Preference-Based Self-Distillation: Beyond KL Matching via Reward Regularization}.
\newblock
\showeprint[arxiv]{2605.05040}~[cs.LG]
\urldef\tempurl%
\url{https://arxiv.org/abs/2605.05040}
\showURL{%
\tempurl}


\bibitem[Yue et~al\mbox{.}(2023)]%
        {MMMU}
\bibfield{author}{\bibinfo{person}{Xiangyu Yue}, \bibinfo{person}{Yu Zheng}, \bibinfo{person}{Zhang Zhang}, \bibinfo{person}{Steven Gao}, \bibinfo{person}{Yuhang Wang}, \bibinfo{person}{Runzhe Chen}, \bibinfo{person}{Yukun Jia}, \bibinfo{person}{Yitong Sun}, \bibinfo{person}{Yizhi Gao}, \bibinfo{person}{Mark Zhao}, {et~al\mbox{.}}} \bibinfo{year}{2023}\natexlab{}.
\newblock \showarticletitle{MMMU: A Massive Multi-discipline Multimodal Understanding and Reasoning Benchmark for Expert AGI}.
\newblock \bibinfo{journal}{\emph{arXiv preprint arXiv:2311.16502}} (\bibinfo{year}{2023}).
\newblock


\bibitem[Zhang et~al\mbox{.}(2026a)]%
        {Embarrassingly}
\bibfield{author}{\bibinfo{person}{Ruixiang Zhang}, \bibinfo{person}{Richard~He Bai}, \bibinfo{person}{Huangjie Zheng}, \bibinfo{person}{Navdeep Jaitly}, \bibinfo{person}{Ronan Collobert}, {and} \bibinfo{person}{Yizhe Zhang}.} \bibinfo{year}{2026}\natexlab{a}.
\newblock \bibinfo{title}{Embarrassingly Simple Self-Distillation Improves Code Generation}.
\newblock
\showeprint[arxiv]{2604.01193}~[cs.CL]
\urldef\tempurl%
\url{https://arxiv.org/abs/2604.01193}
\showURL{%
\tempurl}


\bibitem[Zhang et~al\mbox{.}(2026b)]%
        {OPSDL}
\bibfield{author}{\bibinfo{person}{Xinsen Zhang}, \bibinfo{person}{Zhenkai Ding}, \bibinfo{person}{Tianjun Pan}, \bibinfo{person}{Run Yang}, \bibinfo{person}{Chun Kang}, \bibinfo{person}{Xue Xiong}, {and} \bibinfo{person}{Jingnan Gu}.} \bibinfo{year}{2026}\natexlab{b}.
\newblock \bibinfo{title}{OPSDL: On-Policy Self-Distillation for Long-Context Language Models}.
\newblock
\showeprint[arxiv]{2604.17535}~[cs.CL]
\urldef\tempurl%
\url{https://arxiv.org/abs/2604.17535}
\showURL{%
\tempurl}


\bibitem[Zhang et~al\mbox{.}(2026c)]%
        {Learnwhereto}
\bibfield{author}{\bibinfo{person}{Yan Zhang}, \bibinfo{person}{Daiqing Wu}, \bibinfo{person}{Huawen Shen}, \bibinfo{person}{Can Ma}, {and} \bibinfo{person}{Yu Zhou}.} \bibinfo{year}{2026}\natexlab{c}.
\newblock \bibinfo{title}{Learn where to Click from Yourself: On-Policy Self-Distillation for GUI Grounding}.
\newblock
\showeprint[arxiv]{2605.00642}~[cs.AI]
\urldef\tempurl%
\url{https://arxiv.org/abs/2605.00642}
\showURL{%
\tempurl}


\bibitem[Zhang et~al\mbox{.}(2026d)]%
        {Pi-Play}
\bibfield{author}{\bibinfo{person}{Yaocheng Zhang}, \bibinfo{person}{Yuanheng Zhu}, \bibinfo{person}{Wenyue Chong}, \bibinfo{person}{Songjun Tu}, \bibinfo{person}{Qichao Zhang}, \bibinfo{person}{Jiajun Chai}, \bibinfo{person}{Xiaohan Wang}, \bibinfo{person}{Wei Lin}, \bibinfo{person}{Guojun Yin}, {and} \bibinfo{person}{Dongbin Zhao}.} \bibinfo{year}{2026}\natexlab{d}.
\newblock \bibinfo{title}{$\pi$-Play: Multi-Agent Self-Play via Privileged Self-Distillation without External Data}.
\newblock
\showeprint[arxiv]{2604.14054}~[cs.LG]
\urldef\tempurl%
\url{https://arxiv.org/abs/2604.14054}
\showURL{%
\tempurl}


\bibitem[Zhao et~al\mbox{.}(2026b)]%
        {Self-Distilled}
\bibfield{author}{\bibinfo{person}{Siyan Zhao}, \bibinfo{person}{Zhihui Xie}, \bibinfo{person}{Mengchen Liu}, \bibinfo{person}{Jing Huang}, \bibinfo{person}{Guan Pang}, \bibinfo{person}{Feiyu Chen}, {and} \bibinfo{person}{Aditya Grover}.} \bibinfo{year}{2026}\natexlab{b}.
\newblock \showarticletitle{Self-Distilled Reasoner: On-Policy Self-Distillation for Large Language Models}.
\newblock \bibinfo{journal}{\emph{arXiv preprint arXiv:2601.18734}} (\bibinfo{year}{2026}).
\newblock


\bibitem[Zhao et~al\mbox{.}(2026a)]%
        {Training}
\bibfield{author}{\bibinfo{person}{Zhengyang Zhao}, \bibinfo{person}{Lu Ma}, {and} \bibinfo{person}{Wentao Zhang}.} \bibinfo{year}{2026}\natexlab{a}.
\newblock \bibinfo{title}{Training with Harnesses: On-Policy Harness Self-Distillation for Complex Reasoning}.
\newblock
\showeprint[arxiv]{2605.08741}~[cs.CL]
\urldef\tempurl%
\url{https://arxiv.org/abs/2605.08741}
\showURL{%
\tempurl}


\bibitem[Zheng et~al\mbox{.}(025a)]%
        {zheng2025group}
\bibfield{author}{\bibinfo{person}{Chujie Zheng}, \bibinfo{person}{Shixuan Liu}, \bibinfo{person}{Mingze Li}, \bibinfo{person}{Xiong-Hui Cheng}, \bibinfo{person}{Bowen Yu}, \bibinfo{person}{Chang Gao}, \bibinfo{person}{Kai Dang}, \bibinfo{person}{Yuqiong Liu}, \bibinfo{person}{Kaiwen Men}, \bibinfo{person}{Kejuan Yang}, \bibinfo{person}{Shudan Zhang}, \bibinfo{person}{Xiang Deng}, \bibinfo{person}{Yu Su}, \bibinfo{person}{Huan Sun}, \bibinfo{person}{Minlie Huang}, \bibinfo{person}{Yuxiao Dong}, {and} \bibinfo{person}{Jie Tang}.} \bibinfo{year}{2025a}\natexlab{}.
\newblock \showarticletitle{Group sequence policy optimization}.
\newblock \bibinfo{journal}{\emph{arXiv preprint arXiv:2507.18071}} (\bibinfo{year}{2025a}).
\newblock


\bibitem[Zhou et~al\mbox{.}(2023)]%
        {sft2}
\bibfield{author}{\bibinfo{person}{Chunting Zhou}, \bibinfo{person}{Pengfei Liu}, \bibinfo{person}{Puxin Xu}, \bibinfo{person}{Srini Iyer}, \bibinfo{person}{Jiao Sun}, \bibinfo{person}{Yuning Mao}, \bibinfo{person}{Xuezhe Ma}, \bibinfo{person}{Avia Efrat}, \bibinfo{person}{Ping Yu}, \bibinfo{person}{Lili Yu}, \bibinfo{person}{Susan Zhang}, \bibinfo{person}{Gargi Ghosh}, \bibinfo{person}{Mike Lewis}, \bibinfo{person}{Luke Zettlemoyer}, {and} \bibinfo{person}{Omer Levy}.} \bibinfo{year}{2023}\natexlab{}.
\newblock \bibinfo{title}{LIMA: Less Is More for Alignment}.
\newblock
\showeprint[arxiv]{2305.11206}~[cs.CL]
\urldef\tempurl%
\url{https://arxiv.org/abs/2305.11206}
\showURL{%
\tempurl}


\end{thebibliography}

\appendix

\newpage

\end{document}